\documentclass[11pt,a4paper]{article}
\usepackage{graphicx}
\usepackage{amsmath}
\usepackage{amssymb}
\usepackage{cite}
\usepackage{verbatim} 
\usepackage[usenames]{xcolor}

\newcommand{\bEq}{\begin{equation}}
\newcommand{\eEq}{\end{equation}}
\newcommand{\bEQ}[1]{\begin{equation} \begin{array}{#1}}
\newcommand{\eEQ}{\end{array} \end{equation}}

\newcounter{CorrectThis}

\begin{document}

\pagestyle{empty}

\null

\vfill

\begin{center}

{\Large  
{\bf 
Covariant formula for the driving force  
\\
\vskip 0.2cm
for interface migration
}
}\\

\vskip 1.0cm
A. Morawiec
\vskip 0.2cm 
{Institute of Metallurgy and Materials Science, 
Polish Academy of Sciences, \\ Krak{\'o}w, Poland.
}
\\
E-mail: nmmorawi@cyf-kr.edu.pl \\
Tel.: ++48--122952854, \ \ \  Fax: ++48--122952804 \\

 \end{center}

\vfill

\noindent
{\bf Abstract}
\\
Interfaces of crystalline materials are strongly affected
by the anisotropy of interface energy. 
A key quantity in this context is the interface stiffness tensor, 
which characterizes the response of the interface energy 
to changes in interface orientation 
and plays a role in determining the driving force for interface migration.
The latter is expressible via contraction of the stiffness tensor and 
interface curvature tensor.
Practical computations involving tensors 
necessarily rely on their individual components.
In this work, an explicit component-wise expression 
is derived for the contraction 
of the stiffness and curvature tensors, valid in arbitrary coordinate systems. 
Within this formulation, to get the driving force, 
the curvature tensor is contracted with the pullback 
of a stiffness-related tensor -- originally defined in three-dimensional 
Euclidean space -- onto the interface manifold.
This covariant treatment establishes a physically consistent foundation for three-dimensional computational models 
involving interface stiffness.

\vskip 0.5cm


\vskip 0.5cm

\noindent
\hfill \today

\newpage

\pagestyle{plain}

\section{Introduction}

Interfaces, including free surfaces and grain boundaries,
play a fundamental role in determining the behavior of 
crystalline materials.
Such interfaces are rarely isotropic. 
Their energy, mobility, and other physical properties depend 
strongly on crystallographic orientation, since the atomic packing 
and bonding across an interface change with the orientation 
of the bounding crystal planes (for surfaces) or with the misorientation 
and boundary-plane orientation between two grains (for grain boundaries).
This anisotropy governs microstructural evolution. Understanding 
interface anisotropy 
has therefore become one 
of the important objectives of materials science.

The theoretical foundations of this field trace back to Gibbs,
who introduced the concept of surface free energy as a function 
of orientation. 
Wulff formalized this idea with 
a geometric construction for predicting equilibrium crystal shapes 
from a polar plot of surface energy versus orientation \cite{Wulff_1901}, 
which is a cornerstone of crystal-growth theory.
The rigorous thermodynamic
framework for anisotropic interface equilibrium was established 
by Herring \cite{Herring_1951,Herring_1951a};
he incorporated rotational torque terms 
into interface equilibrium conditions,
and demonstrated that energy variation with orientation creates 
forces resisting boundary rotation. 
Hoffman and Cahn subsequently 
developed the capillarity vector formalism, 
providing a unified framework 
for anisotropic surface and grain boundary thermodynamics
\cite{Hoffman_1972}.

Studies of anisotropic interface behavior rely on a combination 
of experimental and numerical methods. 
Experimental characterization evolved from early empirical measurements 
of thermal grooving at triple junctions and equilibrium facet shapes.
For much of the twentieth century, however, 
the role of boundary-plane orientation remained largely 
unexplored due to the experimental difficulty 
of measuring it directly.
A methodological shift occurred with the advent of 
modern orientation-mapping techniques, 
particularly electron backscatter diffraction 
and three-dimensional reconstruction methods. 
These techniques allowed boundary-plane orientations to be measured 
for large sets of boundaries, enabling the determination 
of grain boundary distributions and boundary energies 
as functions of both misorientation and boundary-plane normal \cite{Saylor_2003a,Saylor_2003b}.
Computational methods have become equally important. 
Crystallographic and thermodynamic models have been used to predict equilibrium shapes and interface stability.  
Atomistic simulations, including density functional theory
and molecular dynamics methods, 
have enabled the calculation 
of interfacial energies from atomic interactions 
(e.g., \cite{Olmsted_2009,Ratanaphan_2015,Bulatov_2014,Sarochawikasit_2021,
Chirayutthanasak_2022}). 
In parallel, simulation methods that explicitly incorporate 
anisotropic interface properties have emerged as versatile 
frameworks for modeling complex microstructural evolution;
see, e.g., 
\cite{Kobayashi_1993,Eggleston_2001,Chen_2002,Tourret_2022,
	Sethian_1996,Bernacki_2011,Bernacki_2023,   Kawasaki_1989,Lepinoux_2010}.

Understanding interfacial anisotropy carries significant implications 
for the broader field of materials science.
Interface anisotropy is a factor in
crystal growth, grain growth, recrystallization, 
crystallographic texture evolution, 
phase transformations, sintering, solidification, and coarsening. 
It is key for phenomena such as faceting, dendritic growth, 
and microstructural pattern formation. 
By controlling crystal growth,
surface energy anisotropy governs thin-film and 
nanomaterial synthesis habits, 
island nucleation modes, catalytic reactivity across specific facets, 
and thermal stability in thin films and nanoparticles.
Studies of interface anisotropy 
link atomic-scale crystallography directly 
to the macroscopic performance of materials and 
provide insights that guide the development of 
materials with tailored properties.

%Anisotropy of interface energy influences properties 
%of crystalline materials. 

The resistance or propensity of an interface 
to changes in its orientation is referred to as interface stiffness.
Mathematically, the interface stiffness is represented by 
a second-rank tensor based on the 
interface energy and its second-order derivatives with respect to 
interface orientation parameters.
Some studies of grain-boundary motion have adopted such tensorial forms, 
defining the stiffness tensor in terms of covariant derivatives 
of the interface energy on the unit sphere of interface normals.

Notably, three related works \cite{Abdeljawad_2018,Moore_2021,Xu_2026}, 
published in recent years, deal with estimating the boundary stiffness tensor. 
The authors 
used atomistic simulation data to develop analytical 
energy models for 
$\Sigma 3$, $5$, $7$, $9$, and $11$ grain boundaries in Ni. 
They draw conclusions on the 
magnitude and anisotropy of interface stiffness compared to that of energy. 
Moreover, they discuss 
the impact of boundary stiffness on the driving force for boundary migration. 
In \cite{Xu_2026}, the estimated stiffness-based driving force 
is compared to observed grain boundary migration rates.
However, an examination of the approach introduced in \cite{Abdeljawad_2018} 
and extended in \cite{Moore_2021,Xu_2026} reveals a conceptual
inconsistency that merits clarification.
While the expressions presented in \cite{Abdeljawad_2018,Moore_2021,Xu_2026} 
appear to be covariant, their implementation effectively reduces 
to a non-tensorial approach based on principal curvature directions.

Like in the differential geometry of surfaces, 
tensor formalism is generally superior to using principal 
curvature directions. It provides 
a systematic and comprehensive framework 
that handles complex geometric behaviors.
Employing tensors ensures physical consistency; 
the coordinate-independent approach acts as a built-in 
check, helping to avoid physically unsound results.
Tensor notation enables straightforward, unambiguous transformations 
between reference frames and provides concise expressions 
that reduce intricate geometric relationships 
to a compact mathematical form.

The covariant formulation of the driving force 
for interface migration is of significant importance 
in materials science, particularly for grain-growth modeling. 
First, an explicit formula is more straightforward to interpret 
and implement than a descriptive scheme 
based on principal curvature directions; 
it eliminates the need to numerically search for principal directions 
on the interface and align external data 
with that local geometry. 
Second, explicit expressions are directly applicable 
to analytically defined functions, 
including both analytical approximations of 
experimental or simulation interface data 
and purely theoretical functions used for model validation.

This paper revisits the formulation of the driving force for interface migration 
as the contraction of the interface stiffness and curvature tensors
from a covariant perspective.
Building on results originally noted in a different context \cite{Morawiec_2000}, 
an explicit component-wise expression for the stiffness–curvature contraction 
is described.
It is based on differential geometry and valid 
for arbitrary coordinate systems on both the interface 
and the unit sphere on which the energy function is defined. 
The formulation
naturally involves the pullback of a stiffness-related tensor 
from Euclidean space onto the interface manifold. 
The resulting expressions clarify the geometric basis 
of the interface stiffness tensor and 
provide a mathematically consistent 
and practically usable foundation for computational treatments of interface stiffness.

The results of this work are applicable 
to various three-dimensional interface modeling methods. 
Computational models for simulating anisotropic grain growth 
need constitutive relations defining 
the driving force for boundary migration. 
For sharp-interface methods -- such as vertex or level-set models -- 
an explicit formula for the driving force 
is provided for direct implementation. 
For the diffuse-interface approach, such as the phase-field model, 
the results establish the precise target expression 
that should be recovered in the sharp-interface limit. 
Furthermore, the derived formulas and three-dimensional 
analytical solutions can serve as useful benchmarks; 
they allow for verification of numerical implementations 
before proceeding to more complex grain growth simulations.

The remainder of this paper is organized as follows. 
Section \ref{sec_prel} presents the necessary preliminaries. 
Section \ref{sec_ist} recalls the driving force for interface 
migration as a quantity obtained by contracting the 
curvature tensor with the interface stiffness tensor. 
Section \ref{sec_GK} provides a fully covariant 
expression for the driving force. 
Its use is illustrated by simple, analytical examples. 
Some technical aspects of stiffness determination 
are clarified in section \ref{sec_Abdeljawad}. 
The paper concludes with a summary and final remarks.

\section{Preliminaries \label{sec_prel}}

The discussion here is concerned  exclusively with continuous, 
sharp interfaces. 
This restriction entails a set of physical assumptions 
under which atomic-scale details can be neglected, 
and the interface may be treated as a smooth surface. 
While the papers \cite{Abdeljawad_2018,Moore_2021,Xu_2026} 
focus on grain boundaries, the considerations 
presented below apply to a broader class of crystalline interfaces, 
including interfaces between distinct crystalline phases 
as well as solid-liquid interfaces.
In the case of a grain boundary, the misorientation between 
the adjoining crystals is assumed to be fixed;  physically, this 
corresponds to an island grain embedded in 
a matrix of different orientation.
All crystals involved are assumed to be centrosymmetric.

The interface energy density 
$\gamma$ is assumed to depend only on the direction of 
the interface normal $\mathbf{n}$, i.e., 
$\gamma=\gamma(\mathbf{n})$.
It is further assumed that 
$\gamma$ and all other functions are differentiable as many times as required for the analysis.
To avoid complications, 
$\gamma$ is assumed not to exhibit missing interface orientations, 
i.e., it does not lead to faceting; equivalently, the spherical plot of 
$1/\gamma$ is assumed to be convex.

\begin{table}
	\centering
	\renewcommand{\arraystretch}{1.2} 
	\begin{tabular}{rl}
		\hline
		\textbf{Symbol} & \textbf{Description} \\
		\hline
		$\mathbf{e}_i$ & 
		Unit vector along the $i$-th axis of the crystal-fixed system \\
		$n^i$, \, $\mathbf{n} = n^i \mathbf{e}_i$ & 
		Unit vector normal to the interface \\
		$s^i$ & Unnormalized vector normal to the interface  \\
		$\vartheta^{\mu}$ $(\mu=1,2)$ & 
		General coordinates of $\mathbf{n}$ on the unit sphere \\
		$\theta, \psi$ & 
		Spherical coordinates of $\mathbf{n}$ \\
		$a_{\mu\nu}$ & 
		Metric tensor on the unit sphere \\[.25cm] 
		$x^i = x^i(u^\alpha)$ & 
		Cartesian coordinates of a point on the interface \\        
		$u^{\alpha}$ $(\alpha=1,2)$ & 
		Local coordinates on the interface \\
		$g_{\alpha\beta}$ & 
		Interface metric tensor  \\	
		$g$ & 
		Determinant of the interface metric tensor \\
		$H$ & 
		Interface mean curvature \\
		$\mathbf{K}$, $K_{\alpha\beta} = - b_{\alpha\beta}$ & 
		Interface curvature tensor  \\[.25cm] 	
		$\gamma = \gamma(\mathbf{n})$ & 
		Interface energy density \\
		$\xi^i$ & Capillarity vector \\
		$\mathbf{\Gamma}(\mathbf{n})$, \, $\mathbf{\Gamma}$,  \, $\Gamma_{\mu\nu}$ &
		Interface stiffness tensor \\		
		$h_{ij}$ & 
		Tensor incorporating energy, its gradient and interface stiffness \\
		$Q_{ij}^{\mu}$, $T_{ij}^{\mu\nu}$ & 
		Coefficients relating  covariant derivatives of energy to $h_{ij}$ \\
		$\widetilde{\Gamma}_{\mu\nu}$ & 
		Effective stiffness tensor \\
		$A_\alpha^{\ \mu}$ & 
		Geometric transformation coefficients \\
		$B_i^{\mu}$   & 
		Coefficients used for defining  
		$Q_{ij}^{\mu}$, $T_{ij}^{\mu\nu}$, $A_\alpha^{\ \mu}$ \\[.25cm] 	
		$f$ & 
		Driving force for interface migration \\
		\hline
	\end{tabular}
	\caption{Glossary of key symbols.}
	\label{table:1}
\end{table}

Summation convention is used throughout the paper.  
The symbols $\delta$ and $\varepsilon$ denote the Kronecker delta and  
permutation symbol, respectively.
The notation $\mbox{diag}(e_1,e_2, \ldots, e_k)$
refers to the square matrix with the entries $e_1,e_2, \ldots, e_k$
on the diagonal and zeros elsewhere. 
To avoid additional symbols,
some tensors are identified by their components.
This article builds upon the formalisms of  
\cite{Abdeljawad_2018,Moore_2021,Xu_2026} and \cite{Morawiec_2000}, and consequently adopts 
some of their notation. To maintain clarity, 
a summary of the key symbols is provided in Table \ref{table:1}.

Three smooth manifolds with distinct coordinate systems 
are considered here.
The first is the Euclidean space of the crystal.
Cartesian coordinates associated with the crystal are denoted by Latin indices. For example, the components of the unit vector 
$\mathbf{n}$ normal to the interface are written as $n^i$ ($i=1,2,3$).
The other two manifolds are the interface surface 
parameterized by coordinates $u^{\alpha}$ ($\alpha=1,2$)
and the unit sphere parameterized by $\vartheta^{\mu}$ ($\mu=1,2$).
In both cases, coordinates and  tensor components
are labeled by Greek indices.
For clarity, quantities 
with indices $\alpha$ or $\beta$ are defined with respect to
the coordinate system on the interface, 
whereas those with indices $\mu$ or $\nu$ are defined with respect to
the coordinate system on the unit sphere.
Differentiation on the interface surface with respect to 
$u^{\alpha}$
is indicated by a dot (e.g., 
$\dot{n}^i_\alpha=\partial n^i/\partial u^{\alpha}$),
whereas differentiation on the unit sphere with respect to 
$\vartheta^{\mu}$ is denoted by 
$\partial_{\mu}$ 
(e.g., $\partial_{\mu} \gamma = \partial \gamma/\partial \vartheta^{\mu}$).
The covariant derivative on the interface surface is marked by
a semicolon subscript `$;$',
whereas the covariant derivative 
on the unit sphere is denoted by $\nabla$.
The metric tensor on the interface is denoted by $g_{\alpha\beta}$,
and that on the unit sphere by $a_{\mu\nu}$.
This notation scheme is summarized in Table~\ref{table:2}.

\begin{table}
\centering
\renewcommand{\arraystretch}{1.2} 
	\begin{tabular}{rlccc}
\hline	
\textbf{Manifold} & \textbf{Indices} &  \textbf{Coordinates}  &  \textbf{Derivative}  
& \textbf{Covariant derivative}  \\
\hline		
3D Euclidean space & $i,j,k$ & Cartesian  
&  & \\
2D interface       & $\alpha, \beta$ & arbitrary 
&
${\dot{\mbox{\textcolor{lightgray}{\rule{1.3ex}{2.0ex}}}}\;\!}_\alpha$ & $\ \ \mbox{\textcolor{lightgray}{\rule{1.3ex}{2.0ex}}\:\!}_{;\alpha}$  \\
2D unit sphere     & $\mu, \nu$  & arbitrary  
& 	
$\partial_\mu$	 & $\nabla_\mu$  \\
\hline
	\end{tabular}
\caption{Summary of the notation convention used to distinguish tensors on different manifolds.}
\label{table:2}
\end{table}

\section{Interface stiffness tensor \label{sec_ist}}

The driving force for interface migration 
is affected by the curvature of the interface and its energy.  
In the presence of interface anisotropy, 
this driving force is frequently written in the form
\begin{equation}
	f=\left(\gamma + \frac{\partial^2 \gamma}{\partial \varphi_1^2} \right) \kappa_1+
	\left(\gamma + \frac{\partial^2 \gamma}{\partial \varphi_2^2} \right) \kappa_2 ,
	\label{eq:Herring}
\end{equation}
where 
$\kappa_1$ and $\kappa_2$ are the principal curvatures of the interface and 
the angles $\varphi_1$ and $\varphi_2$
describe changes of the interface normal in planes orthogonal to the principal curvature directions. 
Expression~\eqref{eq:Herring} was originally derived 
by Herring as the chemical potential near a smoothly curved interface
\cite{Herring_1951} and has since been used in theoretical 
and computational studies of interface motion; 
see, e.g., \cite{Lobkovsky_2004,Blixt_2022,Florez_2022,Yang_2025}.

A more general formula for the driving force employs tensor notation and 
is written as
\begin{equation}
	f=\mathbf{K} : \mathbf{\Gamma} \ ,
	\label{eq:Abdel}
\end{equation}
where 
$\mathbf{K}$ 
is the curvature tensor of the interface, 
$\mathbf{\Gamma}$ is the interface stiffness tensor, 
and the double dot product denotes the contraction 
operation \cite{Gurtin_1988}.
This formal representation of $f$ is a covariant  
generalization of Herring’s result.
Its physical meaning is intended to be preserved 
under changes of coordinate system. 
In computational practice, however, ensuring covariance 
requires the use of appropriate tensor components.
In component notation, eq.~\eqref{eq:Abdel} may have the form
$f=K^{\alpha \beta} \widetilde{\Gamma}_{\alpha \beta}$.

In \cite{Abdeljawad_2018}, 
the interface stiffness tensor $\mathbf{\Gamma}$ is defined 
as $\mathbf{\Gamma} = \gamma \mathbf{I}  
+ \nabla_{\mathbf{n}} \nabla_{\mathbf{n}} \gamma$, 
where $\mathbf{I} $ is the `identity tensor'
and $\nabla_{\mathbf{n}} \nabla_{\mathbf{n}} \gamma$ stands for the Hessian of $\gamma$ on the unit sphere. 
(See also \cite{Du_2007}.)
Its components 
$\Gamma_{\mu \nu}$
are specified with respect to coordinates on the unit sphere --
the domain of the energy function $\gamma$.
The components $\Gamma_{\mu \nu}$ are stated to be based on 
the covariant derivative of $\gamma$,
and  the impression is conveyed that its contraction with the 
curvature tensor is a covariant form of~\eqref{eq:Herring}. 
However, a closer examination reveals conceptual difficulties in this formulation. 
The tensors $K^{\alpha\beta}$ and $\Gamma_{\mu\nu}$ 
are situated on different manifolds and with respect to different metrics.
The curvature tensor is on the interface 
manifold, whereas the stiffness tensor 
is on the unit sphere 
of interface normals. 
A contraction of $K^{\alpha\beta}$ and $\Gamma_{\mu\nu}$
is therefore not well defined unless 
one of the tensors is first transformed appropriately. 
Indices of $K^{\alpha\beta}$ are manipulated
using the interface metric $g_{\alpha\beta}$, 
whereas index operations on $\Gamma_{\mu\nu}$  
are performed using the metric $a_{\mu\nu}$ on the unit sphere.
The inconsistency becomes explicit in the isotropic case:
contracting $\Gamma_{\mu\nu}=a_{\mu\nu}\check{\gamma}$ with $K^{\alpha\beta}$
does not, in general, yield the required result
$f = K^{\alpha\beta} g_{\alpha\beta}\check{\gamma} 
= K^{\alpha}_{\ \alpha}\check{\gamma} = 2H\check{\gamma}$,
where $H$ is the mean curvature of the interface.\footnote{If 
the mixed-component tensors $K^{\alpha}_{\ \beta}$ and $\Gamma^{\mu}_{\ \nu}$
are used, the direct contraction produces the correct result in the isotropic case
($\Gamma^{\mu}_{\ \nu} = \delta^{\mu}_{\ \nu} \check{\gamma}$), but
it fails for the anisotropic part of the stiffness tensor. 
This is because the stiffness tensor is modified by the metric of the sphere 
$a^{\mu\nu}$ ($\Gamma^{\mu}_{\ \nu} = a^{\mu\kappa} \Gamma_{\kappa\nu}$), 
whereas the curvature tensor is modified by the metric of the interface
$g_{\alpha\beta}$ ($K^{\alpha}_{\ \beta} =  K^{\alpha\lambda} g_{\lambda\beta}$).}

By definition, the curvature tensor components $K^{\alpha\beta}$ are given 
with respect to coordinates $u^\alpha$ on the interface surface 
and are intrinsically tied to the surface geometry. 
These coordinates determine not only the curvature but also the local normal 
vector $\mathbf{n}$. In contrast, the coordinates 
$\vartheta^\mu$ parameterize the orientation of $\mathbf n$ 
alone and have no direct connection to the surface geometry 
from which $K^{\alpha\beta}$ is derived.
Consequently,  $f$ 
is a function of position on the interface
(i.e., it depends on $u^\alpha$)
and cannot be represented as a function of $\mathbf{n}$
or $\vartheta^\mu$.
The $\mathbf{n}$-dependent functions of type $\mathbf{\Gamma}(\mathbf{n}):\mathbf{K}$ 
in \cite{Xu_2025_PhD,Xu_2026}
represent quantities related to the stiffness-based driving force, 
rather than the driving force itself.

Although the works \cite{Abdeljawad_2018,Moore_2021,Xu_2026} 
invoke the covariant representation of eq.~\eqref{eq:Abdel}, 
they effectively abandon the tensorial formulation 
in favor of an approach based on principal curvature directions, 
resulting in non-tensorial forms similar to eq.~\eqref{eq:Herring}.
While a procedure based on principal curvatures, when properly implemented, gives 
results quantitatively identical to the covariant 
formula, the former is algebraically more complex.
Moving into a local coordinate system of principal directions 
requires diagonalizing a matrix at every single point 
on the interface, whereas a direct covariant contraction 
avoids that eigenvalue problem.
Furthermore, determining the driving force requires 
aligning the energy-based stiffness data -- 
which is external to the interface geometry -- 
with the local principal directions.

\section{Covariant expressions for $\mathbf{K}:\mathbf{\Gamma}$ \label{sec_GK}}

The question arises as to how $\mathbf{K}:\mathbf{\Gamma}$ 
is expressed explicitly in terms of tensor components. 
Establishing such an expression is 
essential for practical computations involving interface stiffness, 
as numerical implementations of tensorial relationships
require evaluation of tensor components.
In light of the methodology used in 
\cite{Abdeljawad_2018,Moore_2021,Xu_2026}, it is therefore timely 
to revisit this issue and present a proper
formulation in a manner accessible to researchers working on interface and
grain-boundary phenomena.

While it is possible that an explicit component-based 
representation of $\mathbf{K}:\mathbf{\Gamma}$ exists in 
the broader literature on theoretical mechanics, mathematical physics, 
or applied differential geometry, it does not appear to be 
cited or used in practice-oriented studies of interface migration.
A covariant expression for the second-order covariant derivative tensor 
contracted with the curvature tensor can be found, 
in a different context, in an appendix to \cite{Morawiec_2000}. 
There, the result appears only incidentally 
within comments regarding the capillarity vector, 
and the concept of interface stiffness is not explicitly mentioned.
Consequently, the relevance of this expression to contemporary 
treatments of interface stiffness has gone unnoticed.

This section provides an explicit covariant expression 
for $f = \mathbf{K} : \mathbf{\Gamma}$, 
outlines steps needed to compute $f$, 
and presents several illustrative examples. 
To interpret the formula introduced in \cite{Morawiec_2000}, several standard concepts from the differential geometry of surfaces must first be recalled.

\subsection{Relevant elements of differential geometry of surfaces }

Let  $x^i = x^i(u^\alpha)$
denote the Cartesian coordinates of a point on the surface
parametrized by local coordinates $u^\alpha$. 
With a dot indicating partial differentiation with respect 
to $u^\alpha$,
the metric tensor
of the surface is given by
$$
g_{\alpha\beta} = \dot{x}^i_{\alpha}\dot{x}^i_{\beta} \ .
$$
The contravariant metric tensor $g^{\alpha\beta}$ is defined by 
$g_{\alpha\kappa} g^{\kappa\beta} = \delta_\alpha^{\ \beta}$.
The vector normal to the surface, defined as 
the vector product of the tangent vectors $\dot{x}^i_1$ and $\dot{x}^i_2$, is 
$$
s^i = \frac{1}{2}\,\varepsilon_{ijk}\varepsilon^{\alpha\beta}
\dot{x}^j_\alpha \dot{x}^k_\beta 
$$
and satisfies 
$s^i s^i = \det[g_{\alpha\beta}] = g$. 
The unit vector normal to the surface is 
$n^i = s^i/\sqrt{g}$.
The curvature tensor
of the surface is defined as
$$
b_{\alpha\beta} = -\dot{x}^i_\alpha \dot{n}^i_\beta \ .
$$
The tensor $b_{\alpha\beta}$ used in \cite{Morawiec_2000} 
follows the sign convention associated with an 
outward-pointing normal
and is the negative of the curvature tensor $\mathbf{K}$ of 
eq.~\eqref{eq:Abdel}, i.e.,   
$$
b_{\alpha \beta} = - K_{\alpha \beta} \ . 
$$
The trace $K^\alpha_{\ \alpha}$ yields twice the mean curvature,
$K^\alpha_{\ \alpha} = 2H$.
Through the Weingarten equations,
\begin{equation}
\dot{n}^i_\alpha = - b_\alpha^{\ \beta} \dot{x}^i_\beta \ ,  
\label{eq_Weingarten}
\end{equation}
the derivatives of the normal vector are expressed 
in terms of derivatives of the position vector.
The Gauss map of a surface maps each point on the surface 
to its normal, i.e., one has $n^i=n^i(u^\alpha)$.
For more details, see, e.g., \cite{Goetz_1970}.

\subsection{Effective stiffness components}

Like Herring’s eq.~\eqref{eq:Herring}, the note in \cite{Morawiec_2000}
concerns surfaces that minimize the total energy of the system. The approach is
based on the Euler--Lagrange equations applied to the Lagrangian
$L(x^i,\dot{x}^i_{\alpha}) = \gamma \sqrt{g} - \lambda x^i s^i$,
and makes use of the parameter-invariance conditions.
The generalized force acting on a surface element,
$$
E^i =
\frac{\mathrm{d}\ }{\mathrm{d} u^{\alpha}}
\frac{\partial L}{\partial \dot{x}^i_{\alpha}}
- \frac{\partial L}{\partial x^i} \ ,
$$
contains a term involving the contraction of the stiffness-related tensor with the
curvature tensor.
This term was first expressed in \cite{Morawiec_2000} as
\begin{equation}
	f = g^{\alpha\beta}\dot{\xi}^i_\alpha \dot{x}^i_\beta \ ,
	\label{eq_Mor_1}
\end{equation}
where 
\begin{equation}
	\xi^i = \frac{\partial (\sqrt{g} \, \widetilde{\gamma} )}{\partial s^i} \ ,
	\label{eq_capillarity_vec}
\end{equation}
is the Hoffman--Cahn capillarity vector \cite{Hoffman_1972}, 
and the energy density $\widetilde{\gamma}$ 
is a function of $s^i$: 
$\widetilde{\gamma}(\mathbf{s}) = 
\gamma(\mathbf{s}/\sqrt{\mathbf{s} \cdot \mathbf{s}}) = \gamma(\mathbf{n})$.
From now on, the notation will be relaxed for simplicity,
and the tilde over $\gamma$ will be omitted.
As noted in \cite{Morawiec_2000}, 
eq.~\eqref{eq_Mor_1} can be equivalently written in the form
\begin{equation}
	f = - b^{\alpha\beta}\dot{x}^i_\alpha \dot{x}^j_\beta h_{ij} \ ,
	\label{eq_Mor_2}
\end{equation}
where
\begin{equation}
h_{ij} = \gamma \delta_{ij} +
g \frac{\partial^2\gamma}{\partial s^i \partial s^j} \ .
\label{eq_h_tensor}
\end{equation}
Given that this expression for $f$ resembles that used in
\cite{Abdeljawad_2018,Moore_2021,Xu_2026}, and noting that 
the derivation of eq.~\eqref{eq_Mor_2} is omitted in \cite{Morawiec_2000},
the connection between
eqs.~\eqref{eq_Mor_1} and \eqref{eq_Mor_2} is provided in Appendix~A.

Equation~\eqref{eq_Mor_2} represents the contraction of the curvature tensor 
$K^{\alpha \beta}$
with
the tensor whose components are given by
\begin{equation}
	\widetilde{\Gamma}_{\alpha \beta}
	= \dot{x}^i_{\alpha} \dot{x}^j_{\beta} h_{ij}  \ .
	\label{eq_Mor_3}
\end{equation}
With this $\widetilde{\Gamma}_{\alpha \beta}$,
the formula $f=K^{\alpha\beta} \widetilde{\Gamma}_{\alpha\beta}$ is strictly covariant and valid for any admissible coordinates on the interface.
While $h_{ij}$ is determined by the energy function $\gamma$ alone, 
$\widetilde{\Gamma}_{\alpha \beta}$  
depends on the interface geometry through 
the tangent vectors $\dot{x}^i_{\alpha}$. 
The tensor $\widetilde{\Gamma}_{\alpha \beta}$
is the pullback of  $h_{ij}$ 
from that space to the 
interface. (For a description of the pullback operation, 
see, e.g., \cite{Dubrovin_1984}.)

Similar to $\mathbf{\Gamma}(\mathbf{n})$ considered in \cite{Abdeljawad_2018,Moore_2021,Xu_2026}, $h_{ij}$ 
incorporates information regarding the interface stiffness. 
However, unlike $\mathbf{\Gamma}(\mathbf{n})$, 
which is defined on the unit sphere, $h_{ij}$ is expressed
in the crystal-fixed Cartesian coordinate system of the 
surrounding three-dimensional space, 
and this makes its interpretation less intuitive. 
The primary question is how $h_{ij}$ 
relates to the stiffness defined intrinsically on the sphere. 

To find out, one needs to determine the relationship between 
the second term of $h_{ij}$ and the covariant derivatives of $\gamma$.
It can be shown that $\nabla_{\mu} \gamma$ and 
$\nabla_{\mu} \nabla_{\nu} \gamma$ are related to 
$g \, \partial^2 \gamma/\partial s^i \partial s^j$ via 
a system of six linear algebraic equations; see Appendix~B. 
Solving this system with respect to the latter leads to 
\begin{equation}
	g \, \frac{\partial^2 \gamma}{\partial s^i \partial s^j}
	=
	Q_{ij}^{\mu} \nabla_{\mu} \gamma
	+
	T_{ij}^{\mu\nu} \, \nabla_{\mu} \nabla_{\nu} \gamma \ ,
	\label{eq:dsds_Q_T}
\end{equation}
where 
the coefficients $Q_{ij}^{\mu}$ and $T_{ij}^{\mu\nu}$ are given by 
\begin{equation}
	Q_{ij}^{\mu} = - 
	\left(n^i  B_j^{\mu} + n^j  B_i^{\mu} \right) 	
	\ , \ \mbox{  } \ \ 
	T_{ij}^{\mu\nu} = 
	\left(B_i^{\mu}  B_j^{\nu} + B_i^{\nu}  B_j^{\mu} \right)/2 \ ,
	\label{eq:T_formula}
\end{equation}
and 
$$
B_i^{\mu} =  a^{\mu \nu} \partial_{\nu} n^i \ .
$$
The forms of the coefficients  $B_i^{\mu}$, $Q_{ij}^{\mu}$ and $T_{ij}^{\mu\nu}$
depend solely on the choice of the coordinates $\vartheta^{\mu}$ on the 
unit sphere. 
In particular, for spherical coordinates
$(\vartheta^1,\vartheta^2)=(\theta,\phi)$ 
with 
$\mathbf{n}=[\cos\phi \sin\theta \  \sin\phi \sin\theta \  \cos\theta ]$,
the  metric tensor is 
$[a_{\mu\nu}] = \mbox{diag}(1,\sin^2\theta)$, 
and the explicit forms of 
$B_i^{\mu}$ 
are
$$
\begin{array}{lll}
	B_{1}^{1} = \cos \theta \cos \phi \ , & 	
	B_{2}^{1} = \cos \theta \sin \phi \ , &
	B_{3}^{1} = - \sin \theta \ , \\
	
	B_{1}^{2} = -\csc\theta \sin \phi \ , \ \ & 			
	B_{2}^{2} = \csc \theta \cos \phi \ , \ \ & 
	B_{3}^{2} = 0 \ .
\end{array}
$$
The coefficients $Q_{ij}^{\mu}$ and $T_{ij}^{\mu\nu}$ are symmetric 
($Q_{ij}^{\mu}=Q_{ji}^{\mu}$ and
$T_{ij}^{\mu\nu}=T_{ij}^{\nu\mu}=T_{ji}^{\mu\nu}$),
and since $n^i B_i^{\mu} =  0$  
and $B_i^{\mu} B_i^{\nu}=a^{\mu\nu}$, they 
satisfy the identities
$\delta^{ij} T_{ij}^{\mu\nu} = a^{\mu\nu}$,
$a_{\mu\nu} T_{ij}^{\mu\nu}  = \delta_{ij} - n^i n^j$ and 
$\delta^{ij} Q_{ij}^{\mu} = 0$.
Moreover,
with $\dot{x}^i_\alpha$ orthogonal to $n^i$, one has
$\dot{x}^i_\alpha \dot{x}^j_\beta \, Q_{ij}^{\mu} = 0$
and
$\dot{x}^i_\alpha \dot{x}^j_\beta \, T_{ij}^{\mu\nu} a_{\mu\nu} = 
\dot{x}^i_\alpha \dot{x}^j_\beta \,  \delta_{ij}$.
Consequently, one obtains 
\begin{equation}
	\widetilde{\Gamma}_{\alpha\beta}=	
	\dot{x}^i_\alpha \dot{x}^j_\beta T_{ij}^{\mu\nu} \, \Gamma_{\mu\nu}  	
	\label{eq_Mor_4}
\end{equation}
with 
\begin{equation}
\Gamma_{\mu\nu} = \gamma a_{\mu\nu} + \nabla_{\mu} \nabla_{\nu} \gamma \ , 
\label{eq_Mor_Gamma_s}
\end{equation}
and 
$a_{\mu\nu}$, $T_{ij}^{\mu\nu}$ and $\nabla_{\mu} \nabla_{\nu} \gamma$ 
computed in the coordinates $\vartheta^{\mu}$ on the sphere and 
ultimately expressed in \eqref{eq_Mor_4} as functions of
$u^{\alpha}$ through the Gauss map 
$n^i=n^i(u^{\alpha})$ or $\vartheta^{\mu}=\vartheta^{\mu}(u^{\alpha})$.
Equation (\ref{eq_Mor_4}) 
implies that the transformation of 
$\Gamma_{\mu\nu}$ preceding its contraction with the curvature 
tensor has the coefficients 
$\dot{x}^i_\alpha \dot{x}^j_\beta T_{ij}^{\mu\nu}$.
This transformation can be simplified further.

Based on the structure of $T_{ij}^{\mu\nu}$, 
the tensor $\widetilde{\Gamma}_{\alpha\beta}$ can be expressed in 
its final and most natural form 
\begin{equation}
	\widetilde{\Gamma}_{\alpha\beta}=
	A_\alpha^{\ \mu} A_\beta^{\ \nu} \, \Gamma_{\mu\nu}  \ ,
	\label{eq:AA_formula}
\end{equation}
where 
$$
A_\alpha^{\ \mu} = \dot{x}^i_\alpha  B_i^{\mu} \ . 
$$
With the above, the explicit component-based formula for the driving force is 
$$
f=K^{\alpha\beta} A_\alpha^{\ \mu} A_\beta^{\ \nu} \, \Gamma_{\mu\nu}  \ . 
$$
The geometric transformation coefficients $A_\alpha^{\ \mu}$ 
map covariant components from the domain 
of the energy function (the unit sphere) to the interface manifold. 
In particular, mapping the stiffness $\Gamma_{\mu\nu}$ defined on the sphere
onto the interface to obtain the effective stiffness $\widetilde{\Gamma}_{\alpha\beta}$ 
is the covariant equivalent of aligning the
components of $\mathbf{\Gamma}(\mathbf{n})$ with 
local principal curvature directions.

Returning to the tensor $h_{ij}$, its physical significance differs 
from that of the conceptually simpler stiffness tensor $\Gamma_{\mu\nu}$.
The explicit
relation between these tensors follows from 
eqs.~\eqref{eq_h_tensor} and \eqref{eq:dsds_Q_T}
\begin{equation}
h_{ij}=
\gamma n^i n^j + Q_{ij}^{\mu} \nabla_{\mu} \gamma +
T_{ij}^{\mu\nu} \Gamma_{\mu\nu} \ . 
	\label{eq:h_Gamma}
\end{equation}
Thus, $h_{ij}$ is nonzero even if $\Gamma_{\mu\nu}$ vanishes, but  
only the last term affects the driving force $f=\mathbf{K}:\mathbf{\Gamma}$.
The tensor 
$h_{ij}$ carries the combined information of the energy $\gamma$, its gradient $\nabla_{\mu} \gamma$, and the stiffness $\Gamma_{\mu\nu}$. 
It is easy to verify that 
these constituent elements can be extracted through the projections
\begin{equation}
\gamma  =  h_{ij} n^i n^j \ , \ \ \ \ \ \ 
\nabla_{\mu} \gamma =  
-  h_{ij} n^j  \left(\partial_{\mu}  n^i \right) \ , \ \ \ \ \ \ 
\Gamma_{\mu\nu}  =  
h_{ij} \left(\partial_{\mu}  n^i \right) \left(\partial_{\nu} n^j \right) \ . 
	\label{eq:h_Gamma_inv}
\end{equation}
There is an analogy between $h_{ij}$ and the capillarity vector. 
Their defining expressions \eqref{eq_capillarity_vec}, 	
\eqref{eq:capil_2} and \eqref{eq_h_tensor} 
rely on derivatives with respect to the same variables. 
Similarly to \eqref{eq:h_Gamma}, the capillarity vector can be expressed as
$$
\xi^i = \gamma n^i + B_i^{\mu} \nabla_{\mu} \gamma \ .
$$
Its pullback onto the interface, 
$\widetilde{\zeta}_{\alpha} = \dot{x}^i_{\alpha} \xi^i$ (cf. eq.~\eqref{eq_Mor_3}), satisfies $\widetilde{\zeta}_{\alpha} = A_\alpha^{\ \mu} \nabla_{\mu} \gamma$ 
(cf. eq.~\eqref{eq:AA_formula}), and  
the projections $\gamma = \xi^i n^i$ and $\nabla_{\mu} \gamma = \xi^i (\partial_{\mu} n^i)$ correspond to the relations in \eqref{eq:h_Gamma_inv}.
Moreover, the vector $h_{ij} n^j = \gamma n^i - B_i^{\mu} \, \nabla_{\mu} \gamma$ 
is related to the capillarity vector. 
Both are combinations of the normal component $\gamma n^i$ 
and the tangential torque component $B_i^{\mu} \, \nabla_{\mu} \gamma$.
Clearly, the magnitude of $h_{ij} n^j$ is the same as that of 
the capillarity vector, and 
the vector $\gamma n^i$ represents the midpoint 
between $h_{ij} n^j$ and $\xi^i$ in the embedding space.

The expression \eqref{eq:AA_formula} is similar to
the Gauss-map-based pullback  
$
\dot{\vartheta}^{\mu}_{\alpha} \, \dot{\vartheta}^{\nu}_\beta \, 
\Gamma_{\mu\nu}
$ 
of the tensor $\Gamma_{\mu\nu}$ 
from the sphere to the interface.
Moreover, by expressing the interface energy 
$\gamma(\mathbf{n})$
as a function of the interface 
coordinates $u^{\alpha}$ via $n^i=n^i(u^{\alpha})$, 
one may compute the covariant derivative $\gamma_{;\alpha \beta}$ 
with respect to the interface coordinates
and subsequently construct the tensor 
$\gamma g_{\alpha\beta} + \gamma_{;\alpha \beta}$.
Clearly, the  tensors 
$\dot{\vartheta}^\mu_\alpha \, \dot{\vartheta}^\nu_\beta \, \Gamma_{\mu\nu}$ 
and $\gamma g_{\alpha\beta} + \gamma_{;\alpha \beta}$ 
-- both defined on the interface manifold -- 
have physical meanings and numerical values that differ 
from $\widetilde{\Gamma}_{\alpha \beta}$.
A more comprehensive study would be required 
to establish regimes in which
these tensors approximate $\widetilde{\Gamma}_{\alpha \beta}$.

In practice, interfacial energy functions derived from atomistic 
simulations or phenomenological models are frequently expressed 
as continuous, piecewise differentiable functions exhibiting cusps 
at specific orientations. 
At the exact `vertices' of these cusps, the stiffness tensor is undefined. 
However, this does not undermine the utility of the covariant formalism; 
for every other orientation within the open subdomains -- 
including interfaces whose normals approach a cusp vertex -- 
the derived expressions remain valid and applicable.
It is important to recognize that while the singular orientations 
are crystallographically special, 
they form a subset of measure zero within the orientation space.
The expressions
are applicable within smooth domains, while the discrete vertices 
constitute limiting cases. 
In interface modeling, 
the primary physical complexity lies not in the isolated cusps 
vertices themselves, but in the sharp energy gradients surrounding them. 
By accurately handling these smooth regions, 
the covariant approach provides a foundation 
upon which special orientations can be systematically integrated, 
maintaining both mathematical consistency and computational tractability.

\begin{figure}
	\begin{picture}(300,470)(0,0)
		\put(100,270){\resizebox{9.0 cm}{!}{\includegraphics{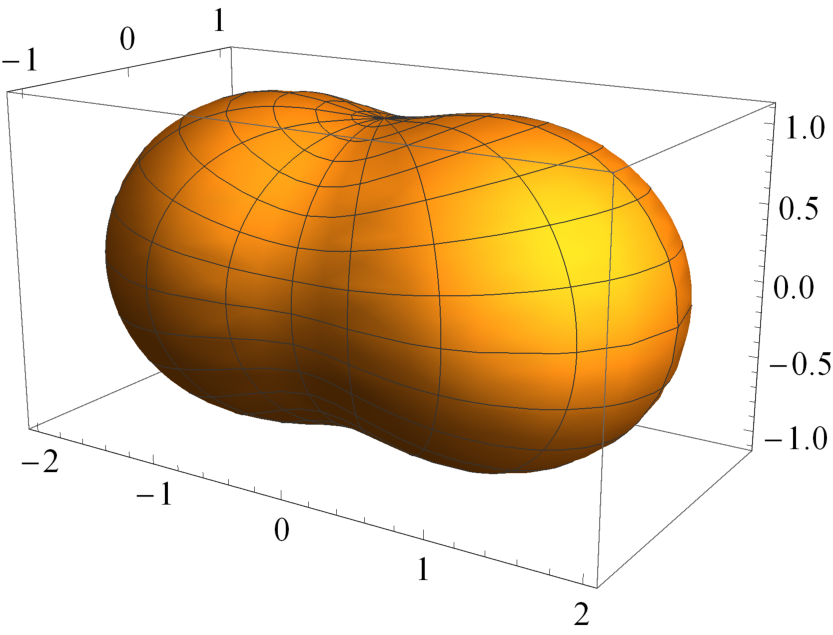}}}
		\put(120,0){\resizebox{7.0 cm}{!}{\includegraphics{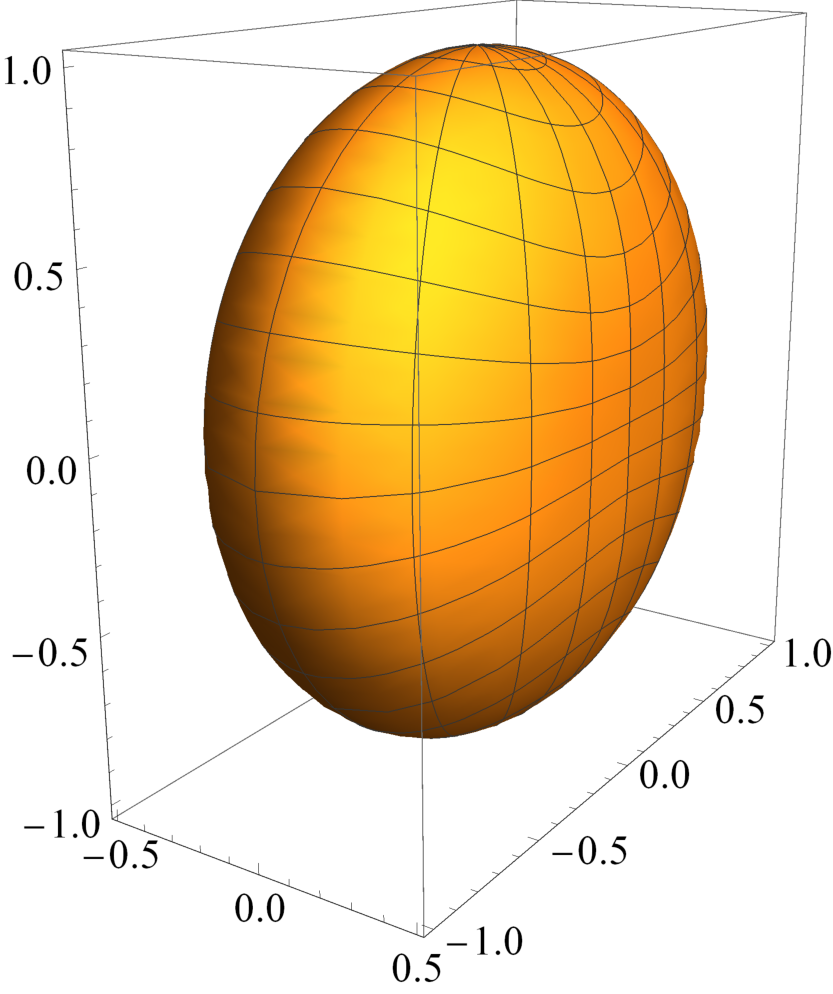}}}	
		\put(100,460){\textit{a}}
		\put(250,273){\small $\gamma n^1$}
		\put(356,412){\small $\gamma n^3$}
		\put(100,250){\textit{b}}
		\put(187,3){\small $n^1/\gamma$}
		\put(106,195){\small $n^3/\gamma$}
	\end{picture}
	\vskip 0.0cm
	\caption{
		Spherical plots of the energy function 
		\eqref{eq:example_energy}  (\textit{a})		
		and
		its inverse $1/\gamma$ (\textit{b}).
		The  $1/\gamma$-plot is nearly planar at $\mathbf{n} = \pm \mathbf{e}_1$.
	}
	\label{Fig_df_energy}
\end{figure}

\begin{figure}
	\begin{picture}(300,470)(0,0)
		\put(100,247){\resizebox{8.0 cm}{!}{\includegraphics{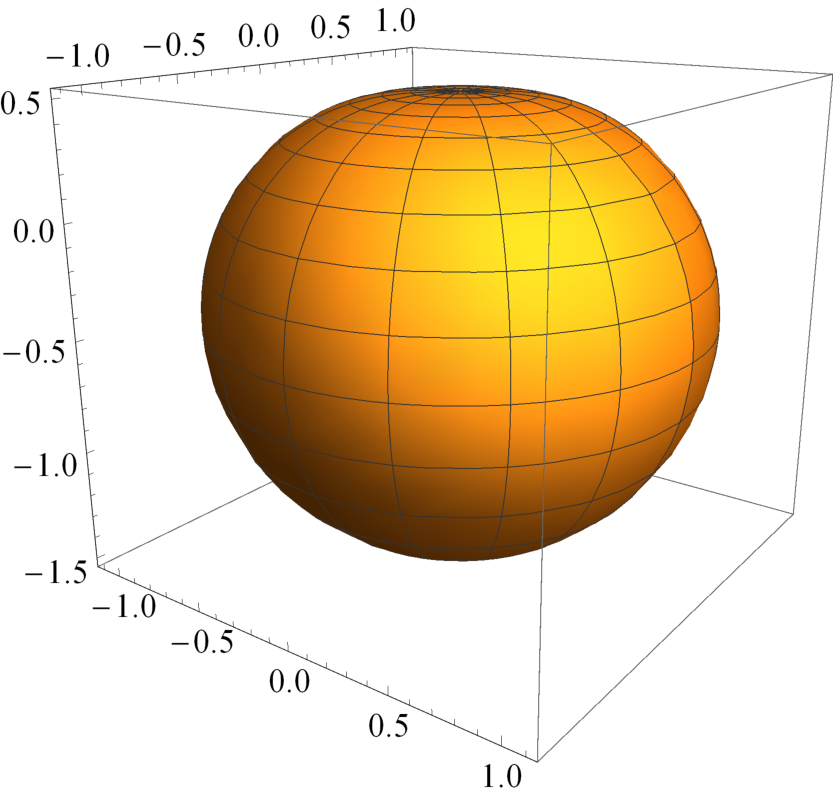}}}
		\put(120,0){\resizebox{8.0 cm}{!}{\includegraphics{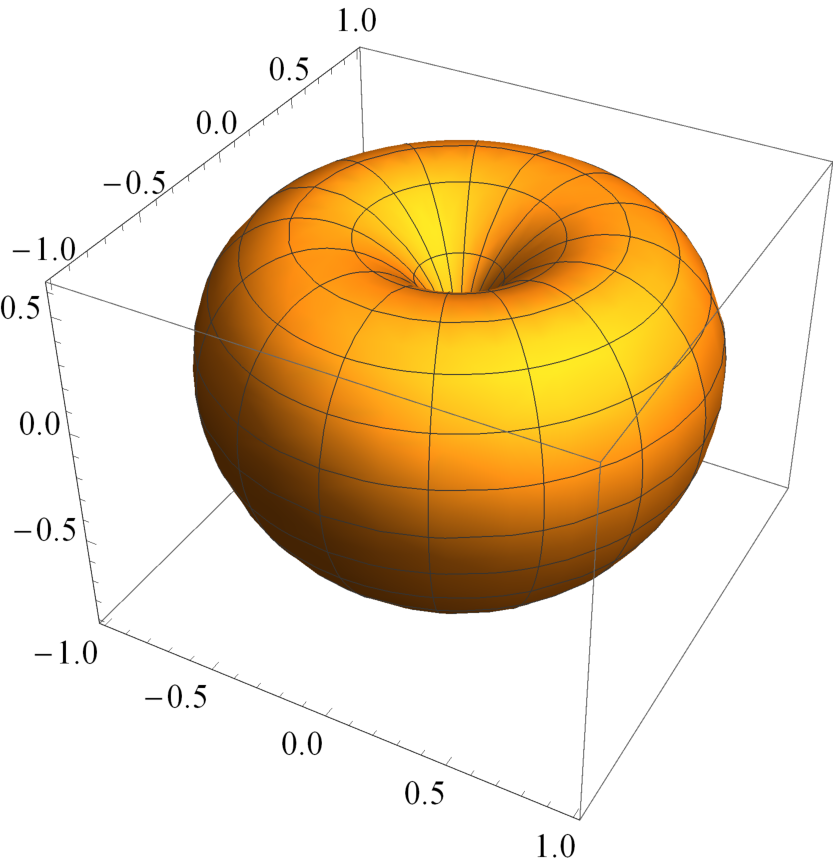}}}		
		\put(90,460){\textit{a}}
		\put(214,250){\small $x^1$}
		\put(92,414){\small $x^3$}
		\put(90,210){\textit{b}}
		\put(228,3){\small $H \mathbf{u} \cdot \mathbf{e}_1$}
		\put(94,134){\small $H \mathbf{u} \cdot \mathbf{e}_3$}
	\end{picture}
	\vskip 0.0cm
	\caption{
		Interface surface defined by eq.~\eqref{eq:example_interface} (\textit{a}). The corresponding spherical plot of the mean curvature 
		is shown in (\textit{b}).
	}
	\label{Fig_df_grain}
\end{figure}

\subsection{A computational procedure \label{sec:comput_steps}}

Since $\widetilde{\Gamma}_{\alpha\beta}$ can be expressed 
in various ways in terms of 
$\dot{x}^i_\alpha$, $A_\alpha^{\ \mu}$, $B_i^{\mu}$, $T_{ij}^{\mu\nu}$ 
and $h_{ij}$, several computational paths involving combinations 
of these quantities may be used to determine the driving force $f$.
Regardless of the specific path, 
the process reduces to computing and contracting a number of 
tensorial quantities.

In the analytical case, 
the input consists of the interface equation $x^i(u^{\alpha})$ 
and the interface energy $\gamma$ 
as a function of the unit normal $n^i(\vartheta^{\mu})$, with 
both $x^i$ and $n^i$ defined with respect to 
a crystal-fixed Cartesian reference system. 
A step-by-step procedure for computing the driving force $f$ can be as follows:
From the interface equation, first derive the tangent vectors $\dot{x}^i_\alpha$, 
the normal vector $n^i$, and the curvature tensor $K^{\alpha\beta}$ 
as functions of the surface coordinates $u^{\alpha}$. 
The specific interface orientation parameters $\vartheta^{\mu}$ 
are then extracted from the normal vector $n^i$ at $u^{\alpha}$, 
followed by the determination of the coefficients $A_\alpha^{\ \mu}$. 
Using the energy function $\gamma(\mathbf{n}(\vartheta^{\mu}))$, 
calculate the stiffness components $\Gamma_{\mu\nu}$ 
and the effective stiffness 
$\widetilde{\Gamma}_{\alpha\beta}=A_\alpha^{\ \mu} A_\beta^{\ \nu} \Gamma_{\mu\nu}$. 
Finally, the driving force $f$ is obtained 
by contracting $\widetilde{\Gamma}_{\alpha\beta}$ with $K^{\alpha\beta}$.

In practice, where interface and energy data are discrete, 
computational schemes can be designed to mimic this analytical workflow. 
As in the approach of \cite{Abdeljawad_2018,Moore_2021,Xu_2026}, 
a continuous energy function can be fitted to the energy data, 
with a similar procedure applied to the interface geometry. 
However, contemporary interface migration simulation methods
employ technique-specific strategies for representing the 
interface and the energy density, 
alongside distinct numerical schemes for calculating derivatives. 
Consequently, the procedures for implementing the formulas for 
$K^{\alpha\beta}$ and $\widetilde{\Gamma}_{\alpha\beta}$
will depend on the simulation method employed.

\subsection{Examples}

The following examples illustrate the use of the presented formulas.
The first case considers a prescribed interface geometry with 
an arbitrary energy function. 
The subsequent two examples concern specific energy functions 
and arbitrary interfaces. 
In the last one, both the interface 
geometry and the energy function are explicitly defined.

A particularly useful case for benchmarking is a spherical interface 
of radius $r$. For this geometry, the curvature satisfies the relation
$K^{\alpha\beta} A_{\alpha}^{\ \mu} A_{\beta}^{\ \nu} = a^{\mu\nu}/r$.
Consequently, the driving force simplifies to
$$
f = a^{\mu\nu} \Gamma_{\mu\nu}/r = (2\gamma + \Delta_s \gamma)/r \ ,
$$
where $\Delta_s = a^{\mu\nu}\nabla_{\mu}\nabla_{\nu}$ 
denotes the Laplace–Beltrami operator on the unit sphere,
and -- 
assuming the center of the interface is at the origin 
of the Cartesian reference system --
in order to get $f$ at $x^i$, the right hand side of the expression 
is calculated at $n^i = x^i/r$.

In the isotropic case, where the interface energy is independent of
orientation, $\gamma(\mathbf{n})=\check{\gamma}=\mathrm{const}$, 
one has $h_{ij}=\check{\gamma} \, \delta_{ij}$,
the tensor \eqref{eq_Mor_3} reduces to
$\widetilde{\Gamma}_{\alpha\beta}=
\dot{x}^i_\alpha \dot{x}^j_\beta \, \check{\gamma} \, \delta_{ij} =
\check{\gamma} \, g_{\alpha\beta}$, 
and the corresponding driving force takes the form
of the Young-Laplace relation
$f= 
2 H \check{\gamma}$.

Another easily tractable case involves an energy function whose $\gamma$-plot 
is a portion of the Herring sphere; it bounds energies of interfaces 
stable with respect to the
formation of hill-and-valley structures \cite{Herring_1951a},
and the corresponding $1/\gamma$-plot is planar. 
Locally, such a function can be written as
\begin{equation}
\gamma(\mathbf{n})= \mathbf{n} \cdot \mathbf{c} = n^i c^i \ , 
\label{eq:gam_n_c}
\end{equation}
where $\mathbf{c}$ is a constant vector
(such that $\mathbf{n} \cdot \mathbf{c}>0$). 
In this case, based on the definition \eqref{eq_h_tensor}, 
the tensor $h_{ij}$ takes the form
\begin{equation}
h_{ij}=3 n^i n^j (n^k c^k)-(n^i c^j+n^j c^i)  \ . 
\label{eq:gam_n_c_h}
\end{equation}
Although it is nonzero, its contraction with two vectors tangent
to the interface (and hence orthogonal to $\mathbf{n}$) vanishes.
Consequently,
$\widetilde{\Gamma}_{\alpha\beta}
= \dot{x}^i_\alpha \dot{x}^j_\beta h_{ij} =  0$,
i.e., the stiffness factor in the driving force
vanishes and $f=0$. 
This result is consistent with the 
nature of interfaces with 
energies lying on the Herring sphere;
if the normals of an interface are 
within the domain of a single Herring sphere, 
the interface will remain stationary, 
regardless of its geometric complexity.

It is instructive to examine the behavior of 
$\widetilde{\Gamma}_{\alpha\beta}$ and $f$ in a case with 
more general characteristics. 
With $\mathbf{e}_i$ denoting the unit vector along $i$-th axis of the Cartesian coordinate system, 
let the energy function be 
\begin{equation}
	\gamma(\mathbf{n}) = 1 + (\mathbf{e}_1 \cdot \mathbf{n})^2 \ .
	\label{eq:example_energy}
\end{equation}
It is a particular form of the Rapini-Papoular energy -- 
a model commonly used to describe the surface anchoring energy of liquid crystals;
see, e.g., \cite{Kleman_2003,Andrienko_2018}.
The function is simple, 
yet it captures essential features of realistic anisotropic interface energies.
The corresponding Wulff plot is shown in Fig.~\ref{Fig_df_energy}\textit{a}.
At $\mathbf{n} = \pm \mathbf{e}_1$, the components $h_{ij}$ have the form
\eqref{eq:gam_n_c_h}, and the corresponding 
$\widetilde{\Gamma}_{\alpha\beta}$ vanish;
see Fig.~\ref{Fig_df_energy}\textit{b}.
Let the interface be described by
\begin{equation}
	\mathbf{x}(\mathbf{u}) = \mathbf{u} 
	\left(1 - \mathbf{u} \cdot \mathbf{e}_3/2\right) \ ,
	\label{eq:example_interface}
\end{equation}
where 
$\mathbf{u}=\left[ \cos u^2 \sin u^1 \  \sin u^2 \sin u^1 \ \cos u^1 \right]$
(Fig.~\ref{Fig_df_grain}\textit{a}).
One has $\lim_{\mathbf{u} \rightarrow \mathbf{e}_3} K_{\alpha\beta}=0$,
i.e., both the mean and Gaussian curvatures vanish at the `north' pole, 
and the interface is locally flat there
(Fig.~\ref{Fig_df_grain}\textit{b}).

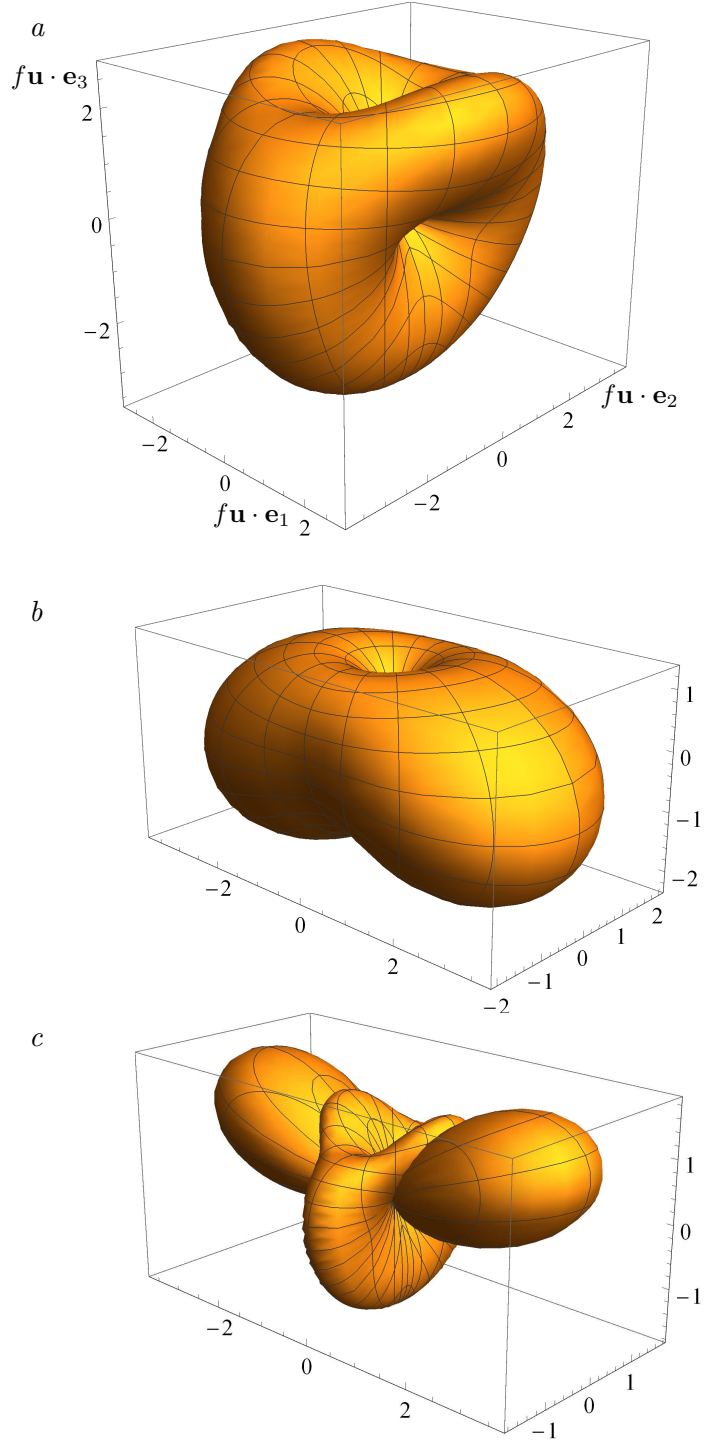
\begin{figure}
	\begin{picture}(300,480)(0,0)
		\put(100,340){\resizebox{7.5 cm}{!}{\includegraphics{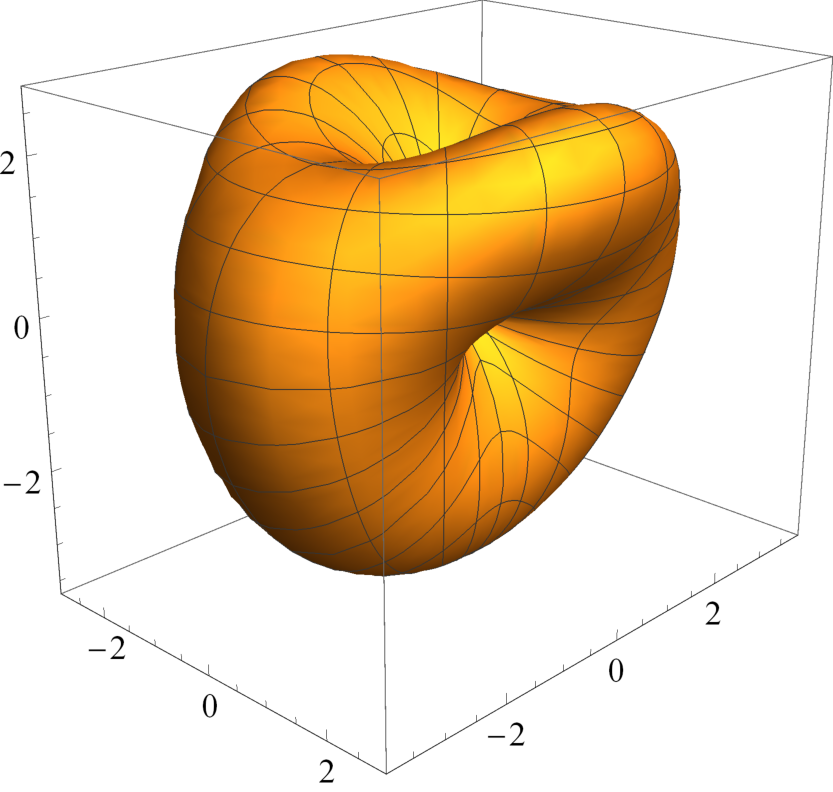}}}
		\put(120,160){\resizebox{7.5 cm}{!}{\includegraphics{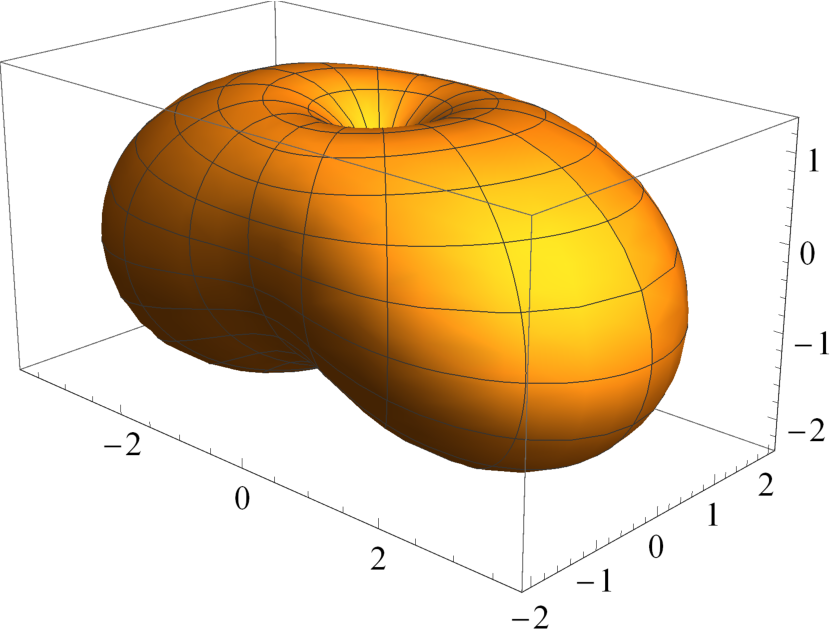}}}
		\put(120,0){\resizebox{7.5 cm}{!}{\includegraphics{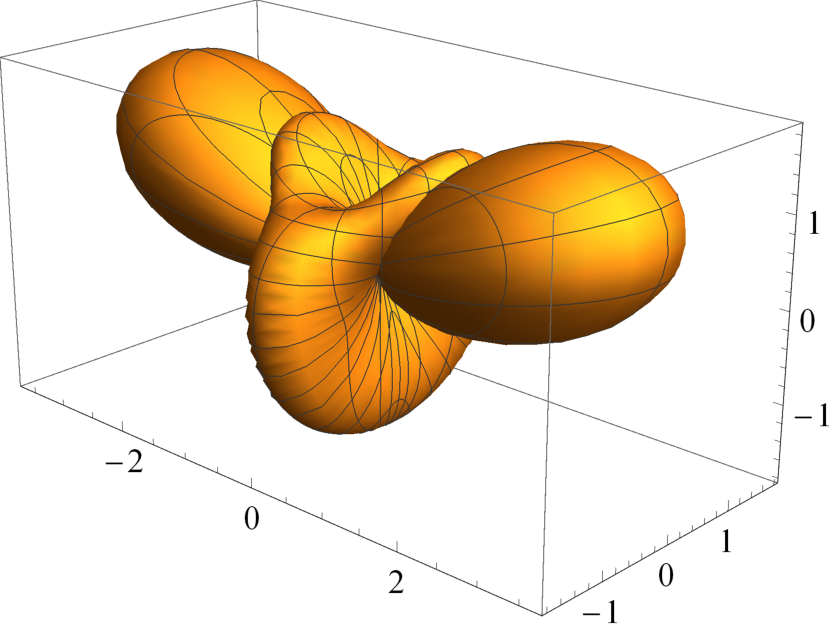}}}
		
		\put(80,530){\textit{a}}
		\put(149,347){\small $f \mathbf{u} \cdot \mathbf{e}_1$}
		\put(295,392){\small $f \mathbf{u} \cdot \mathbf{e}_2$}
		\put(73,512){\small $f \mathbf{u} \cdot \mathbf{e}_3$}
		
		\put(80,310){\textit{b}}
		\put(80,150){\textit{c}}
	\end{picture}
	\vskip 0.0cm
	\caption{
		The driving force $f \mathbf{u}$ calculated based on \eqref{eq_Mor_2}
		for energy function \eqref{eq:example_energy} and the interface \eqref{eq:example_interface}  (\textit{a}). Its isotropic and anisotropic terms are in (\textit{b}) and (\textit{c}), respectively.  
		The coordinate systems in (\textit{b}) and (\textit{c}) are (approximately) aligned with the one in (\textit{a}).
	}
	\label{Fig_df_my_f}
\end{figure}

\begin{figure}[t]
	\begin{picture}(300,360)(0,0)
		\put(80,220){\resizebox{9.0 cm}{!}{\includegraphics{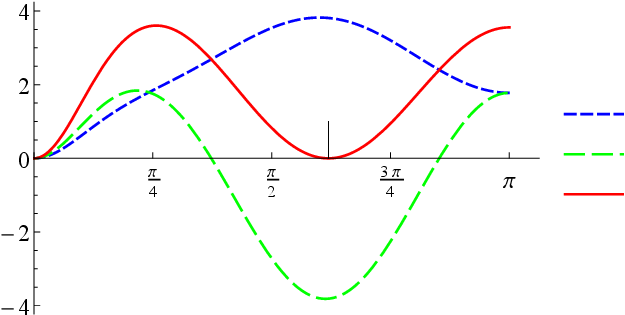}}}
		\put(100,0){\resizebox{7.0 cm}{!}{\includegraphics{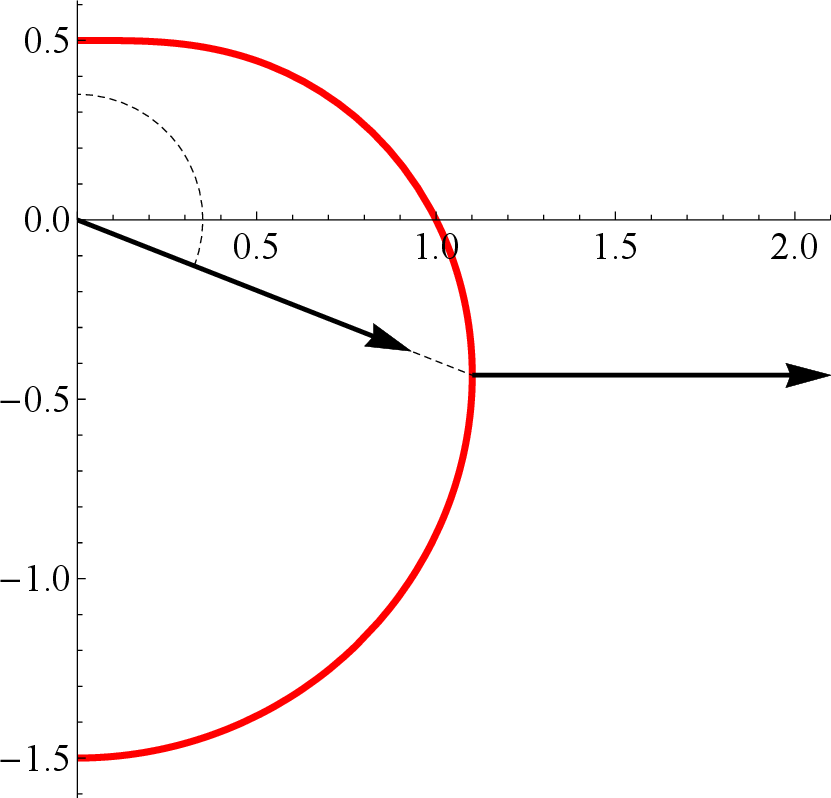}}}
		
		\put(60,350){\textit{a}}
		\put(60,190){\textit{b}}
		\put(302,141){\small $x^1$}
		\put(108,191){\small $x^3$}
		
		\put(340,302){\small isotropic term}
		\put(340,285){\small anisotropic term}
		\put(340,267){\small $f$ based on \eqref{eq_Mor_2}}
		
		\put(84,354){driving force}
		\put(298,274){$u^1$}		
		\put(210,303){$\check{u}^1$}		
		\put(260,345){$u^2=0$}
		\put(128,146){$\check{u}^1$}
		\put(165,110){$\mathbf{u}$}		
		\put(240,105){$\mathbf{n} = \mathbf{e}_1$}		
		
		\put(158,180){\small interface}
		\put(260,170){$u^2=0$}
	\end{picture}
	\vskip 0.0cm
	\caption{
		The driving force $f$ calculated based on \eqref{eq_Mor_2}
		for energy function \eqref{eq:example_energy} and the interface \eqref{eq:example_interface} versus $u^1$ for fixed $u^2=0$ (\textit{a}).
		The angle $\check{u}^1=\arccos \left( \left(1-\sqrt{3}\right)/2\right)$ 
		at which $f$ reaches zero corresponds to the 
		direction $\mathbf{u}$ for which $\mathbf{n}(\mathbf{u}) = \mathbf{e}_1$,
		where $1/\gamma$ is flat and the stiffness tensor vanishes (\textit{b}).  
	}
	\label{Fig_df_my_f_2D}
\end{figure}

With the analytical expressions \eqref{eq:example_energy} and \eqref{eq:example_interface} specified, the driving force 
$f$ can be determined 
using software for symbolic computation
following the procedure outlined in section~\ref{sec:comput_steps}.
Its spherical plot is in Fig.~\ref{Fig_df_my_f}. 
The figure also shows the isotropic and anisotropic energy contributions 
that constitute $f$. 
The function $f$ vanishes at $\mathbf{u} = \mathbf{e}_3$, 
due to the geometry of the interface, 
and also at $\mathbf{u}$ such 
that $\mathbf{n}(\mathbf{u}) = \pm \mathbf{e}_1$, 
reflecting the properties of $\gamma$.
For clarity, sections through the functions shown in Fig.~\ref{Fig_df_my_f} are 
presented in Fig.~\ref{Fig_df_my_f_2D}\textit{a},
for fixed azimuth angle $u^2=0$ and varying polar angle $u^1$. 
The driving force $f$ vanishes 
at the angle $u^1 = \check{u}^1 > \pi/2$, 
where the unit normal $\mathbf{n}$ aligns with $\mathbf{e}_1$;
see Fig.~\ref{Fig_df_my_f_2D}(\textit{b}).
The isotropic contribution is zero only at $\mathbf{u} = \mathbf{e}_3$ 
($u^1 = 0$)
and remains positive elsewhere, whereas the anisotropic contribution 
varies more rapidly and assumes both positive and negative values.

The functions \eqref{eq:example_energy} and \eqref{eq:example_interface} 
may be modified to examine how energy anisotropy and interface curvature affect 
the driving force for migration. 
Clearly, if 
$\gamma(\mathbf{n})$ in \eqref{eq:example_energy}
and
$\mathbf{x}(\mathbf{u})$ in \eqref{eq:example_interface} are scaled 
by $s_{\gamma}$ and $s_{\mathbf{x}}$, respectively, 
$f$ is scaled by $s_{\gamma}/s_{\mathbf{x}}$.
Generally, synthetic functions can be used to draw conclusions 
about the impact of interface stiffness, similar in nature 
to those derived from simulated data in \cite{Abdeljawad_2018,Moore_2021}. 
In the example presented above, 
the maximum isotropic and anisotropic contributions 
are of similar magnitude. 
Other analytically tractable functions can incorporate more realistic features 
while remaining computationally manageable. 
Results from such simple examples demonstrate that 
even with relatively mild energy anisotropy, 
the influence of the anisotropic term can exceed that of the isotropic term.

\section{Methodological aspects of stiffness determination}
\label{sec_Abdeljawad}

Any comparison of the covariant approach to the scheme 
proposed in \cite{Abdeljawad_2018, Moore_2021, Xu_2026} 
would require addressing specific implementation choices 
that affect the results reported in those works.

First, the expression for the Hessian of $\gamma$ 
in \cite{Abdeljawad_2018} diverges from the standard covariant definition.
The second-order covariant derivative, $\nabla_{\mu}\nabla_{\nu}\gamma$, 
of a scalar function $\gamma$ on the unit sphere expressed 
in spherical coordinates is
$$
\left[ \nabla_{\mu} \nabla_{\nu} \gamma \right] =
\left[
\begin{array}{cc}
	\partial_\theta^2  \gamma & 
	\partial_\phi \partial_\theta \gamma-\cot \theta \, \partial_\phi \gamma \\
	\partial_\phi \partial_\theta \gamma-\cot \theta \, \partial_\phi \gamma & 	
	\partial_\phi^2  \gamma +\sin\theta \cos\theta \, \partial_\theta\gamma \\
\end{array}
\right] \ .
$$
Neither these components nor their contravariant 
($a^{\mu\kappa} a^{\nu\lambda} \nabla_{\kappa} \nabla_{\lambda} \gamma$) 
or mixed ($a^{\mu \kappa} \nabla_{\kappa} \nabla_{\nu} \gamma$) forms
coincide with those provided in eq.~(7) of \cite{Abdeljawad_2018}; 
see also the Supplemental Material of \cite{Xu_2026}
and \cite{Xu_2025_PhD}.

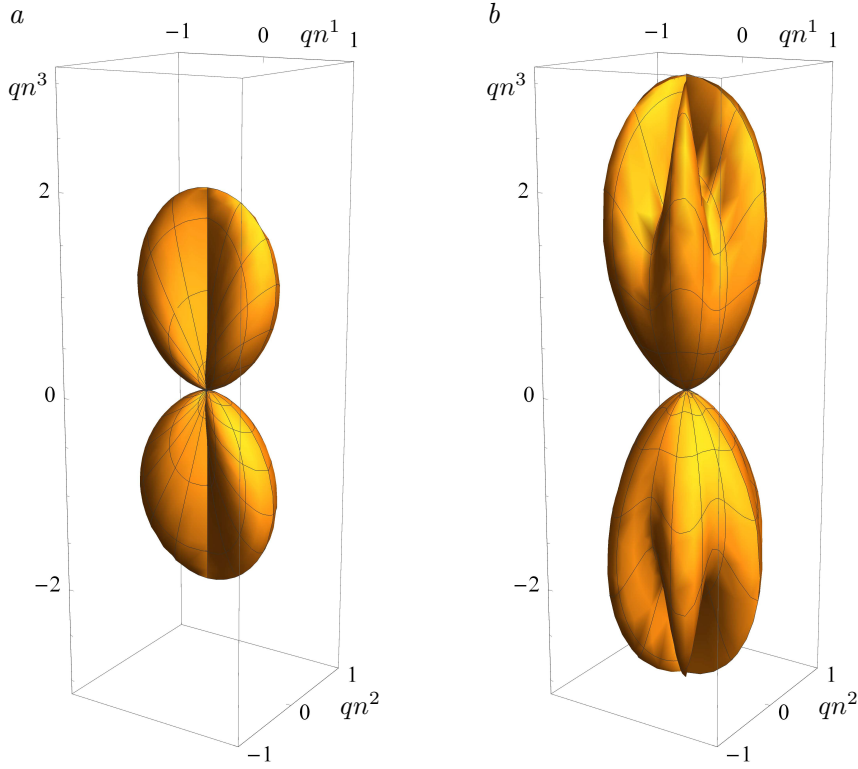
\begin{figure}[t]
	\begin{picture}(300,280)(0,0)
		\put(50,0){\resizebox{4.3 cm}{!}{\includegraphics{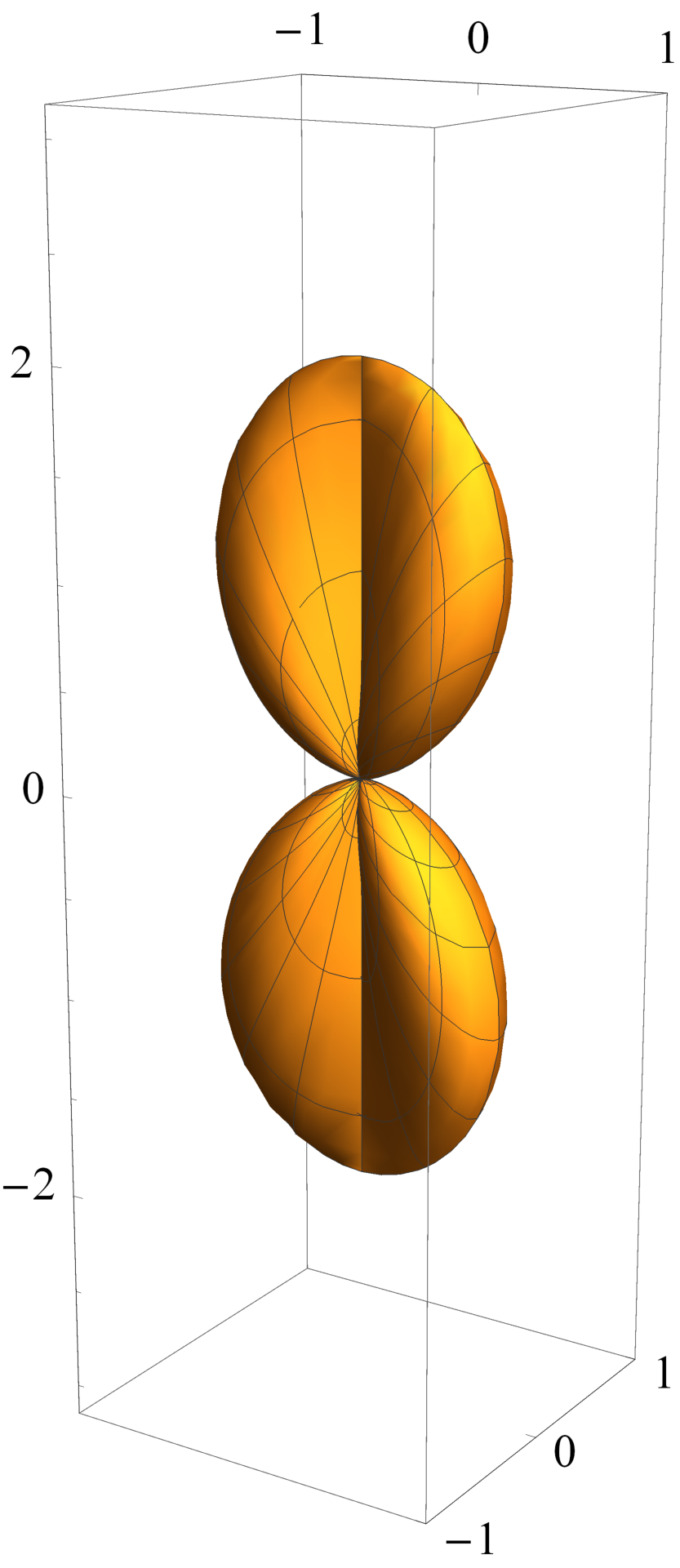}}}
		\put(230,0){\resizebox{4.3 cm}{!}{\includegraphics{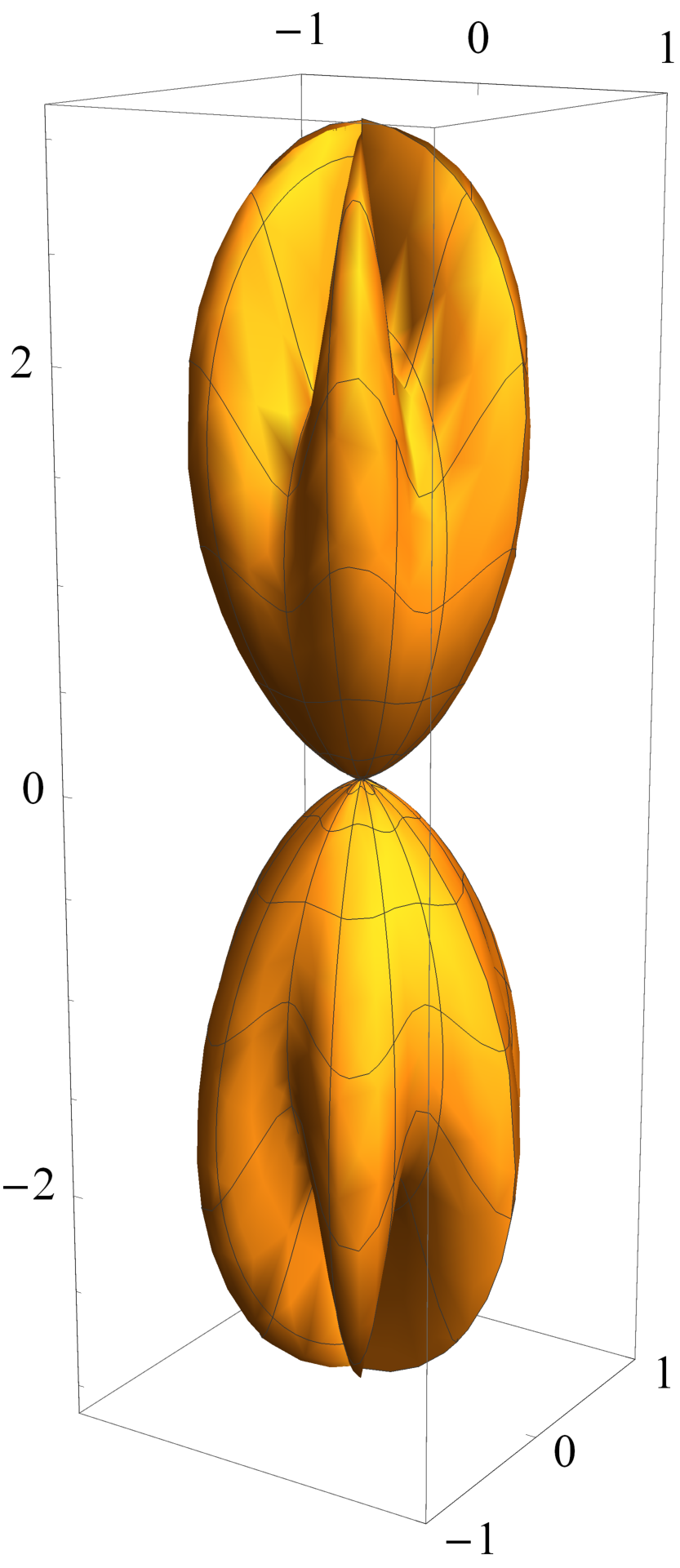}}}
		\put(40,280){\textit{a}}
		\put(150,273){\small $q n^1$}
		\put(165,20){\small $q n^2$}
		\put(40,253){\small $q n^3$}
		\put(220,280){\textit{b}}
		\put(330,273){\small $q n^1$}
		\put(345,20){\small $q n^2$}
		\put(220,253){\small $q n^3$}
	\end{picture}
	\vskip 0.0cm
	\caption{
	Spherical plot of  
	$q =|\Delta_s \gamma - \left(\Delta_s \gamma\right)_{\mbox{\footnotesize approx}}|$ 
	for the energy function $\gamma$ 
	defined in \eqref{eq:example_energy} (\textit{a}) 
	and the energy function from \cite{Abdeljawad_2018} (\textit{b}). 
 Values in (\textit{a}) are dimensionless, while those in (\textit{b}) 
 are given in $\mbox{J}/\mbox{m}^2$.
	}
	\label{Fig_approx}
\end{figure}

Moreover,
the assumption of "smoothly varying energy–inclination profiles" 
\cite{Abdeljawad_2018} does not justify approximating 
$\mathbf{\Gamma}$ by 
$\text{diag}(\gamma +\partial_\theta^2 \gamma, \, \gamma +\partial_\phi^2 \gamma)$. 
The approximation effectively replaces the Hessian
$[\nabla_{\mu} \nabla_{\nu} \gamma]$ given in spherical coordinates 
with  
$[G_{\mu\nu}] =\text{diag}(\partial_\theta^2 \gamma, \, \partial_\phi^2 \gamma)$.
Such an approximation can be poor, even in simple cases. 
For illustration, the scalar 
$\Delta_s \gamma = a^{\mu \nu} \nabla_{\mu} \nabla_{\nu} \gamma$ 
is compared to its approximated counterpart 
$\left(\Delta_s \gamma\right)_{\mbox{\footnotesize approx}} = a^{\mu \nu} G_{\mu\nu}$ 
in Fig.~\ref{Fig_approx} for the function \eqref{eq:example_energy}
and the energy function considered in \cite{Abdeljawad_2018}. 
While no discrepancies occur at the equator ($\theta \approx \pi/2$), 
they become significant at other orientations. 
For example, at $\theta=\pi/4$ and $\phi=0$, 
the values of $\Delta_s \gamma$ and 
$\left(\Delta_s \gamma\right)_{\mbox{\footnotesize approx}}$ 
are $-1$ and $-2$, respectively, 
for the function \eqref{eq:example_energy}, and 
$-0.29 \mbox{J}/\mbox{m}^2$ and $-1.04 \mbox{J}/\mbox{m}^2$ 
for the energy function from \cite{Abdeljawad_2018}. 
Given that the energy functions are known analytically, 
the motivation for using the approximation is unclear.
 
Finally, 
the models fitted to the simulation data
(eqs.~(13) in \cite{Abdeljawad_2018} and eqs.~(7) in \cite{Moore_2021}),
which are linear combinations of products of sine and cosine functions, 
ignore the intrinsic geometry of the sphere.
They are multivalued at the poles 
and vary rapidly with respect to $\phi$ near the 
poles.\footnote{Moreover, to respect symmetries, 
	the $z$-axis is aligned with the principal symmetry axis 
	of a given misorientation. 
	In effect, the energies of boundaries with 
	normals along the symmetry axes are poorly represented.
	For example, 
	in the study of $\Sigma3$ misorientation in Ni \cite{Abdeljawad_2018}, 
	the fitted energy function at the twin boundary takes values between 
	$0.180$ and $0.329\mbox{J}/\mbox{m}^2$, 
	whereas the simulated value is significantly lower 
	at $0.064\mbox{J}/\mbox{m}^2$.
}
The proper approach to modeling smooth functions on the sphere 
is to use spherical harmonics.
%, which form a basis specifically tailored to functions defined on the sphere.
It is well known that estimating second-order derivatives 
of a function by fitting a model to numerical data is inherently difficult. 
The models employed in \cite{Abdeljawad_2018,Moore_2021,Xu_2026} 
do not appear sufficiently refined to reliably estimate 
the stiffness tensor at directions away from the equator.

\section{Concluding remarks}

Energy of crystal interfaces is generally anisotropic, 
meaning it depends on the interface orientation. 
The anisotropy results in a complex behavior, 
where interfaces may resist or exhibit a propensity toward orientation changes.
This property is quantitatively described by an interface stiffness tensor, 
which enters the expression for the driving force governing 
curvature-driven interface motion. 
The determination of the stiffness tensor 
from experimental 
or computed energy data presents a significant challenge.
The tensor involves second order derivatives, 
and numerical differentiation is sensitive to noise;
minor inaccuracies in the data can lead to 
substantial errors in the resulting derivatives.
Drawing conclusions about interface stiffness from specific energy models 
requires significant caution and rigorous validation, 
as similar models may lead to different results. 

On the other hand,
qualitative insights can often be reached without extensive calculation. 
For instance, 
regarding the $\Sigma 3$ boundaries in Ni investigated in \cite{Abdeljawad_2018}, 
the instability of most boundary planes follows from 
a simple inspection of the underlying data; 
as noted in \cite{Morawiec_2025}, 
only a small fraction of data points (44 out of 566) 
lies on the convex hull of the $1/\gamma$-plot,
and only these points represent stable boundaries.
Similarly, instabilities are inherent when the Read–Shockley–Wolf (RSW) 
function is used to model energy cusps for fixed misorientations 
(e.g., \cite{Bulatov_2014});
in such cases, the $1/\gamma$-plot near an RSW-modeled cusp is 
non-convex \cite{Morawiec_2025}, 
which implies interface instabilities in the vicinity of such cusps.
Finally, if the energy model consists 
of patches where the $\gamma$-plots correspond to Herring spheres 
(i.e., the $1/\gamma$-plots are convex hull polyhedra enclosing 
finite sets of data points, cf. \cite{Morawiec_2025}), 
the stiffness is zero everywhere except at the patch boundaries, 
where the derivatives are discontinuous.

Three-dimensional simulations of interface migration 
require the underlying driving force, which is given by  
contraction of the interface curvature tensor and stiffness tensor. 
Practical computations involving tensors rely 
on the manipulation of their individual components. 
Consequently, it is essential to use explicit
expressions that are ready for numerical implementation. 
To ensure the integrity and physical consistency of these implementations, 
it is preferable for the expressions to be covariant, 
as this guarantees that the results remain valid regardless 
of the chosen coordinate system.

In this paper, a fully covariant 
formulation of the contraction of 
interface curvature and stiffness tensors
grounded in differential geometry is described. 
Building on a result previously noted in a different context, 
an explicit component-wise expression is presented; 
this expression is valid for arbitrary coordinate systems 
on both the interface and the unit sphere --  the domain of the energy function.
Within this framework, 
the curvature tensor is contracted with the pullback 
of a tensor defined in three-dimensional 
Euclidean space onto the interface manifold.
It is shown how the tensor resulting from the pullback
can also be obtained by a linear transformation 
of the stiffness tensor defined on the unit sphere. 
This construction naturally incorporates both the anisotropy 
of the interface energy and the local geometry of the surface,
ensuring that the resulting driving force is 
coordinate-independent and physically meaningful.
The covariant formula eliminates the need to 
search for principal directions on the interface 
and align energy data with that local geometry. 
The explicit covariant expressions are directly 
applicable to analytically defined functions.
The component-wise covariant framework 
described above provides a robust foundation 
for future computational studies of anisotropic interface migration.

\vskip 1.0cm

\section*{Appendix A: Derivation of eq.~\eqref{eq_Mor_2}}

The purpose of this appendix is to show that eq.~\eqref{eq_Mor_1}
can be equivalently written in the form~\eqref{eq_Mor_2}.
Since
$$
\frac{\partial \sqrt{g}}{\partial s^j}=
\frac{\partial \sqrt{s^i s^i}}{\partial s^j} = n^j \ , 
$$
the capillarity vector~\eqref{eq_capillarity_vec}
can be expressed as
\begin{equation}
\xi^i = \gamma n^i + \sqrt{g}\,\frac{\partial \gamma}{\partial s^i} \ .
	\label{eq:capil_2}
\end{equation}
Differentiating $\xi^i$ with respect to $u^{\alpha}$
gives 
\begin{equation}
		\dot{\xi}^i_\alpha=
		\dot{\gamma}_\alpha n^i
		+ \gamma \dot{n}^i_\alpha
		\\
		+ n^m \frac{\partial \gamma}{\partial s^i}
		\dot{s}^m_\alpha
		+ \sqrt{g}
		\frac{\partial^2 \gamma}{\partial s^i \partial s^m}
		\dot{s}^m_\alpha
	 \ .
	\label{eq:fderiv03}
\end{equation}
The energy $\gamma$  as a function of $s^i$ is homogeneous of degree zero; hence
$s^j \partial \gamma/\partial s^j = 0$.
Differentiating with respect to $s^i$ results in
\begin{equation}
\frac{\partial \gamma}{\partial s^i}
+ s^j
\frac{\partial^2 \gamma}{\partial s^i \partial s^j} = 0 \ .
	\label{eq:homog01}
\end{equation}
Substituting \eqref{eq:fderiv03} into \eqref{eq_Mor_1} and 
using identity \eqref{eq:homog01}, together with
$\dot{x}^i_\beta n^i = 0$,
leads to
$$
f = g^{\alpha\beta} \dot{x}^i_\beta
\left(
\gamma \dot{n}^i_\alpha
+
\sqrt{g}
(\delta^{jm} - n^j n^m)
\frac{\partial^2 \gamma}{\partial s^i \partial s^j}
\dot{s}^m_\alpha
\right) \ . 
$$
Since 
$\dot{n}^j_\alpha =
(\delta^{jm} - n^j n^m)
\dot{s}^m_\alpha / \sqrt{g}$,
the above expression becomes
$$
f = g^{\alpha\beta} \dot{x}^i_\beta
\left( 
\gamma \dot{n}^i_\alpha +
g \frac{\partial^2 \gamma}{\partial s^i \partial s^j}
\dot{n}^j_\alpha
\right) \ .
$$
Finally, using the Weingarten relations \eqref{eq_Weingarten},
one obtains eq.~\eqref{eq_Mor_2}.

\section*{Appendix B: Expressing $g \, \partial^2 \gamma/\partial s^i \partial s^j$ via covariant derivatives of $\gamma$ on unit sphere}

The objective is to establish a scheme for relating the derivatives
$$
(s^k s^k)\,\frac{\partial^2 \gamma}{\partial s^i \partial s^j}
$$
to the covariant derivatives $\nabla_{\mu} \gamma$ and 
$\nabla_{\mu} \nabla_{\nu} \gamma$ 
of the energy density
$\gamma$, where $\gamma$ is regarded as a function of $s^i$, 
(i.e., $\gamma = \gamma(s^i)$), 
and $s^i$ depend on the coordinates 
$\vartheta^{\mu}$ of $n^i=s^i/\sqrt{s^k s^k}$ on the unit sphere
(i.e., $s^i = s^i(\vartheta^{\mu}))$.
The covariant derivative of
\begin{equation}
\nabla_{\mu} \gamma = \partial_{\mu}\gamma
=
\frac{\partial \gamma}{\partial s^i}\,
\partial_{\mu} s^i 
\label{eq:partialmugamma}
\end{equation}
on the sphere is
$$
\begin{aligned}
\nabla_{\mu} \nabla_{\nu} \gamma 
& =
\partial_{\nu}
\left(
\frac{\partial \gamma}{\partial s^i}\,
\partial_{\mu} s^i
\right)
-
\Gamma^{\kappa}_{\mu\nu}
\left(
\frac{\partial \gamma}{\partial s^i}\,
\partial_{\kappa} s^i
\right)
\\
& = 
\frac{\partial^2 \gamma}{\partial s^i \partial s^j}
(\partial_{\mu} s^i)(\partial_{\nu} s^j)
+
\frac{\partial \gamma}{\partial s^i}\,
\partial_{\mu}\partial_{\nu} s^i
-
\Gamma^{\kappa}_{\mu\nu}
\frac{\partial \gamma}{\partial s^i}\,
\partial_{\kappa} s^i \ ,
\end{aligned}
$$
where $\Gamma^{\kappa}_{\mu\nu}$ denote Christoffel symbols.
Substituting eq.~\eqref{eq:homog01} into this expression yields
$$
\nabla_{\mu} \nabla_{\nu} \gamma 
=
\frac{\partial^2 \gamma}{\partial s^i \partial s^j}
\left(
(\partial_{\mu} s^i)(\partial_{\nu} s^j)
-
s^j \partial_{\mu}\partial_{\nu} s^i
+
s^j \Gamma^{\kappa}_{\mu\nu}\partial_{\kappa} s^i
\right).
$$
Recognizing the covariant second derivative of $s^i$,
$\nabla_{\mu} \nabla_{\nu} s^i=\partial_{\mu}\partial_{\nu} s^i -
\Gamma^{\kappa}_{\mu\nu}\partial_{\kappa} s^i$,
one can write
\begin{equation}
\begin{aligned}
\nabla_{\mu} \nabla_{\nu} \gamma 
	& =
\frac{\partial^2 \gamma}{\partial s^i \partial s^j}
\left(
	(\partial_{\mu} s^i)(\partial_{\nu} s^j)-s^j \, \nabla_{\mu} \nabla_{\nu} s^i
	\right)
\\
& =
\frac{\partial^2 \gamma}{\partial s^i \partial s^j}
\left(
2(\partial_{\mu} s^i)(\partial_{\nu} s^j) -
\nabla_{\mu} \nabla_{\nu} (s^i s^j)/2
\right) \ .
\end{aligned}
\label{eq:auxil}
\end{equation}
Eq.~\eqref{eq:auxil} provides three independent relations for  
$\partial^2 \gamma/\partial s^i \partial s^j$.
Two further relations follow from the homogeneity of $\gamma$; 
combining eq.~\eqref{eq:homog01}
with eq.~\eqref{eq:partialmugamma}, one obtains
\begin{equation}
\partial_{\mu}\gamma +
\frac{\partial^2 \gamma}{\partial s^i \partial s^j}
s^j \partial_{\mu} s^i = 0 \ .
\label{eq:auxil2}
\end{equation}
The final condition also follows from eq.~\eqref{eq:homog01}; one has
\begin{equation}
s^i s^j \frac{\partial^2 \gamma}{\partial s^i \partial s^j} =
- s^i  \frac{\partial \gamma}{\partial s^i} = 0 \ .
\label{eq:auxil3}
\end{equation}
Together, the relations \eqref{eq:auxil}--\eqref{eq:auxil3}
constitute a system of six linear algebraic equations for 
the six independent components of $\partial^2 \gamma/\partial s^i \partial s^j$. 
While solving this system manually 
would be cumbersome, 
it can be accomplished using symbolic computation software.
The resulting solution can be expressed in the form \eqref{eq:dsds_Q_T}.

\clearpage

\bibliographystyle{unsrt}
\bibliography{IntStiffness.bib}

\end{document}